%% file: main.tex
\documentclass[conference]{IEEEtran}
\IEEEoverridecommandlockouts
\usepackage{cite}
\usepackage{amsmath,amssymb,amsfonts}
\usepackage{algorithmic}
\usepackage{graphicx}
\usepackage{textcomp}
\usepackage{xcolor}
\usepackage{pgfplots}
\usepackage{physics}
\usepackage{pgfplotstable}
\usepackage{hyphenat}
\usepackage{arydshln}
\newcommand\blfootnote[1]{%
  \begingroup
  \renewcommand\thefootnote{}\footnote{#1}%
  \addtocounter{footnote}{-1}%
  \endgroup
}

\definecolor{YlOrRd-9-5}{RGB}{253,141,60}
\definecolor{KITpalegreen}{RGB}{130,190,60}
\definecolor{KITcyanblue}{RGB}{80,170,230}
\definecolor{KITlila}{RGB}{163,16,124}

\def\BibTeX{{\rm B\kern-.05em{\sc i\kern-.025em b}\kern-.08em
    T\kern-.1667em\lower.7ex\hbox{E}\kern-.125emX}}
\begin{document}

\title{Low Rate Protograph-Based LDPC Codes for Continuous Variable Quantum Key Distribution}

\author{\IEEEauthorblockN{Kadir G\"um\"u\c s and Laurent Schmalen}
\IEEEauthorblockA{
Karlsruhe Institute of Technology (KIT)\\
Communications Engineering Lab\\
76187 Karlsruhe, Germany\\
\texttt{kadir.guemues@kit.edu}}
}

\maketitle

\begin{abstract}
Error correction plays a major role in the reconciliation of continuous variable quantum key distribution (CV-QKD) and greatly affects the performance of the system. CV-QKD requires error correction codes of extremely low rates and high reconciliation efficiencies. There are only very few code designs available in this ultra low rate regime. In this paper, we introduce a method for designing protograph-based ultra low rate LDPC codes using differential evolution. By proposing type-based protographs, a new way of representing low rate protograph-based LDPC codes, we drastically reduce the complexity of the protograph optimization, which enables us to quickly design codes over a wide range of rates. We show that the codes resulting from our optimization outperform the codes from the literature both in regards to the threshold and in finite-length performance,  validated by Monte-Carlo simulations, showing gains in the regime relevant for CV-QKD.\vspace*{-1ex}
\end{abstract}

\blfootnote{This work has received funding from the European Research Council (ERC) under the European Union’s Horizon 2020 research and innovation programme (grant agreement No. 101001899).}

\section{Introduction}
\input{Sections/Introduction}

\section{LDPC Codes for CV-QKD}
\label{sec:BI}

\subsection{Secret Key Rate}
\input{Sections/SKR}
\subsection{Protograph-based LDPC Codes}
\input{Sections/Protographs}

\subsection{PEXIT Analysis}
\input{Sections/PEXIT}

\section{Optimizing Low Rate Protographs}
\label{sec:OLRP}

\subsection{Protograph Optimization}
\input{Sections/Protograph_Design}

\subsection{Modified PEXIT Analysis}
\input{Sections/Modified_PEXIT}

\section{Results}
\label{sec:Results}

\input{Sections/Results}

\section{Conclusion}
\label{sec:Conclusion}

\input{Sections/Conclusion}


\end{document}

%% file: Sections/Introduction.tex
Low-density parity-check (LDPC) codes are forward error correcting codes first introduced in 1962 \cite{Gallager1962} and then mostly forgotten. After their rediscovery, (e.g., in~\cite{Mackay1996}), LDPC codes have seen wide use in many telecommunication applications because of their performance close to the theoretical limits and the availability of low-complexity decoding algorithms. Most of these applications require codes with high rates, and as such, most research activities have focused on designing high rate codes. However, in recent years, the need for LDPC codes with ultra low rates has increased due to emerging applications such as the reverse reconciliation in continuous variable quantum key distribution (CV-QKD) \cite{Bennett1992}. 

The main performance measure for CV-QKD is the secret key rate (SKR), representing the rate at which secret keys can be exchanged. This SKR depends on the performance of the error correcting code used during reconciliation. In the reconciliation, there is trade-off between the frame error rate (FER) and how close the code operates to capacity. Due to this trade-off, the operating FER of CV-QKD systems is typically very high \cite{Jouguet2011}. 
Therefore, when designing a code for CV-QKD, the error floor behaviour is not the main objective~\cite{Milicevic2017}. It is however important to have codes with low FERs at low SNRs. 

Most previous works on the design of low rate LDPC codes have used multi-edge type (MET) constructions or variants thereof. In \cite[Ch.~7]{Richardson2002}, an MET-LDPC code of rate $\frac{1}{10}$ has been designed by appending a large number of parity check bits to a high rate code akin to a concatenation of an LDPC code with a low-density generator matrix (LDGM) code \cite{Garcia2003}. Additionally, codes of rates down to $
\frac{1}{50}$ were designed based on the rate $\frac{1}{10}$ code of \cite[Ch.~7]{Richardson2002} by further optimizing the degree distributions using differential evolution \cite{Jouguet2011,Storn1997}. More recently, LDPC codes with a cascade structure have been designed down to rates of $\frac{1}{100}$ with thresholds close to capacity~\cite[Ch.~4]{Mani2021}. 

The downside of directly optimizing the degree distributions of MET-LDPC codes is that the optimization is rather complex due to the large amount of degrees of freedom. Furthermore, no simple optimization algorithm exists, as, e.g., for simple irregular LDPC codes. Although these codes perform close to capacity, the range of available rates and designs is sparse. Therefore, with the current methods, designing a specific low rate LDPC code for a particular application is an arduous task. 

One method of reducing the complexity of the code design is by restricting ourselves to protograph-based LDPC codes~\cite{Thorpe2003,Divsalar2009}. Prior work has resulted in codes with good thresholds, while limiting the search space and thus the complexity. However, most protograph-based LDPC code designs have targeted the high rate regime, e.g., \cite{Divsalar2006} or \cite{Uchikawa2014}. An exception is~\cite{Divsalar2005}, presenting a few lower rate protograph-based LDPC codes of rates $\frac{1}{10}$ and above.

The problem with designing low rate protograph-based LDPC codes is that the size of the protograph increases quadratically with decreasing code rate, leading to an exponentially increasing search space and rendering traditional numerical optimization techniques for protographs useless. 
In this paper, we introduce a simplified representation for low rate protograph-based LDPC codes based on partitioning the rows and columns of the protograph into types, called type-based protographs (TBPs). We show that by viewing protographs as TBPs, we can significantly reduce the search space of the optimization, making it possible to design efficient LDPC codes of arbitrarily low rates. We obtain codes using the TBP optimization that outperform codes in the literature, both in threshold and finite-length performance. We verify our results with Monte-Carlo simulations and obtain gains of up to 0.12\,dB  in the regime relevant for CV-QKD. We furthermore propose expanded TBPs, which alleviate some of the drawbacks that this method presents when designing higher rate codes.

%% file: Sections/SKR.tex
 The SKR is the main measure used for evaluating the performance of a CV-QKD system. For conventional CV-QKD protocols, the SKR is given by\cite{Laudenbach2018ContinuousVariableQK}
\begin{equation}
\text{SKR} = (1-\text{FER})(\beta I_{\text{AB}}-\chi_{\text{BE}}).
\label{SKR}
\end{equation} Here, $I_{\text{AB}}$ is the mutual information between Alice (transmitter) and Bob (receiver), $\chi_{\text{BE}}$ is the Holevo information, a measure for the amount of information a potential eavesdropper has on the secret keys, and $\beta$ is the reconciliation efficiency. The reconciliation efficiency is defined as $\beta  = \frac{R}{I_{\text{AB}}}$ and represents the multiplicative gap between $R$, the rate of the code, and $I_{\text{AB}}$. The mutual information $I_{\text{AB}}$ is the upper bound for the code rate, and $\beta \leq 1$ as $R \leq I_{\text{AB}}$. Typically, there is a trade-off between $\beta$ and the FER, which affects the SKR in (\ref{SKR}): the higher $\beta$ becomes, the higher the FER. Because of this trade-off, the operational FER for CV-QKD systems tends to be quite large, which is why the error floor behaviour of a code is less important for CV-QKD. In this paper we assume the use of the one-dimensional reconciliation protocol as described in~\cite{Milicevic2017}. In this protocol, the equivalent channel relevant for the decoder can be modeled as a binary-input additive white Gaussian noise (BI-AWGN) channel. We will focus on this model in the remainder of the paper. 

%% file: Sections/Protographs.tex
\begin{figure}[t!]
    \centering
	\includegraphics[width=1\columnwidth]{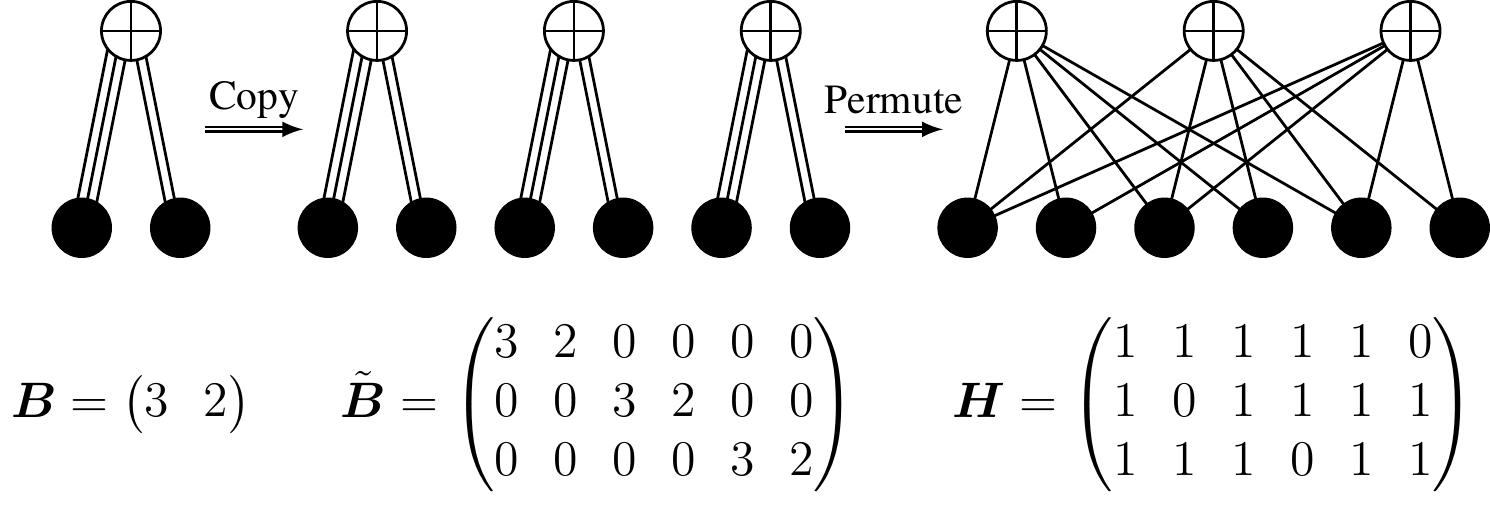}
    \caption{Creating a parity check matrix by lifting a protograph representing a rate $\frac{1}{2}$ LDPC code by a factor 3.}
    \label{fig:Protograph_Lifting}
\end{figure}
Protograph-based LDPC codes are a special case of MET-LDPC codes \cite{Thorpe2003}. Protograph-based codes are constructed from the $q$-cover of the relatively small bipartite graph described by the protomatrix $\boldsymbol{B}$ which conveys the main properties of the code. $b_{i,j}$ denotes the entry of $\boldsymbol{B}$ at row $i$ and column $j$. In contrast to the Tanner graph representation of LDPC codes, the protograph may contain multiple edges. The code itself is constructed by placing $q$ copies of the protograph next to each other (note that these have no interconnecting edges) and permuting the edges between the different copies of the protograph, such that the relation between the group of edges is respected. This operation is called lifting the protograph. Figure \ref{fig:Protograph_Lifting} shows an example of a rate $\frac{1}{2}$ protograph being lifted by a factor $q = 3$.

The rows of a protomatrix represent the check nodes of a code, while the columns represent the variable nodes. A non-zero value $b_{i,j}\neq 0$ indicates that there are edge connections between check node $i$ and  variable node $j$. Every non-zero value in the protograph represents its own edge type. 

\begin{figure}[t!]
    \centering
	\includegraphics[width=1\columnwidth]{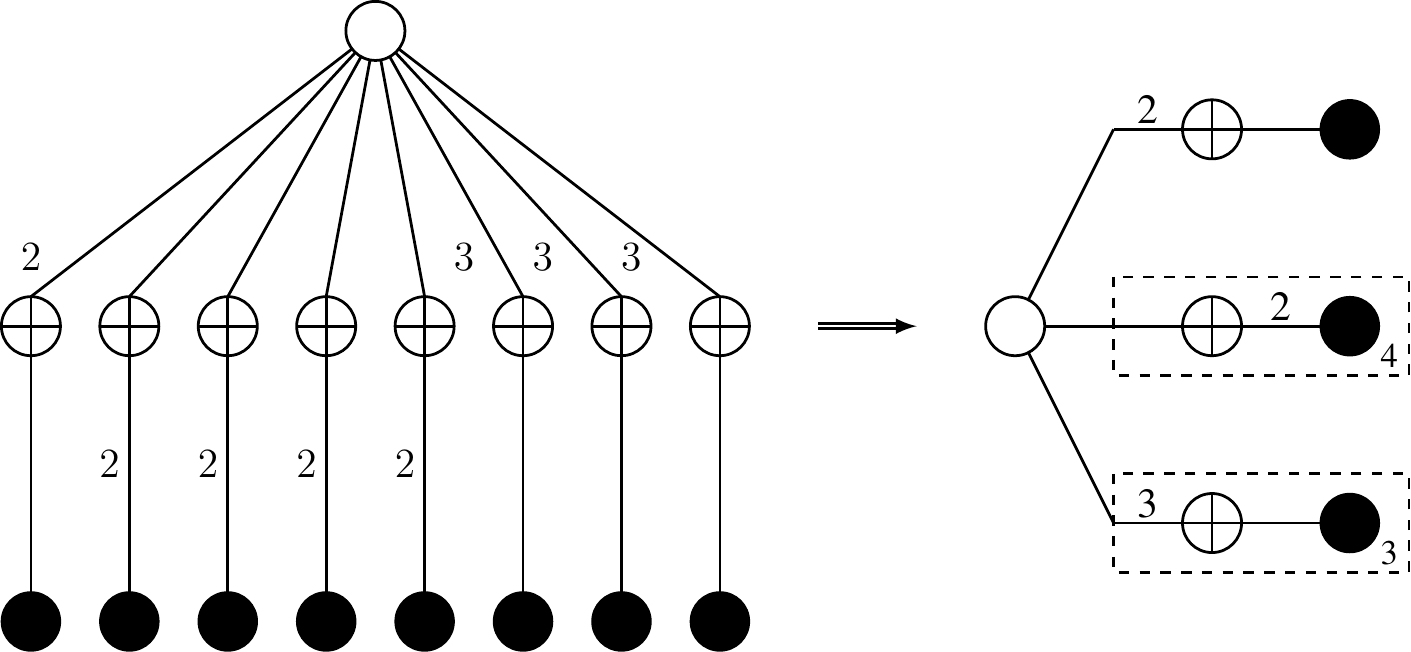}
	\caption{Protograph and its simplified representation for the rate $\frac{1}{8}$ AR4A code~\cite{Divsalar2005}.}
    \label{fig:Protograph_Representation}
\end{figure}

The rate of protograph-based LDPC codes is determined by the dimensions of the protomatrix. Let $m$ and $n$ be the number of rows and columns of the protomatrix, respectively. Additionally, let $n_{\text{p}}$ denote the amount of punctured columns in the protograph. In that case, the design rate of the code is given by $R = \frac{n-m}{n-n_{\text{p}}}$. 
\\\indent Protographs are often visualised by a bipartite graph, giving a simple and elegant representation. However, for low rate protographs this representation becomes very cluttered due to the large amount of variable and check nodes. We simplify this representation by grouping together similar nodes that occur multiple times within the protograph. A dashed box around a subgraph  with a number $\kappa$ in its lower right corner indicates that we repeat this subgraph $\kappa$ times and connect each copy to its originating node. An exemplary protograph for the rate $\frac{1}{8}$ AR4A code from \cite{Divsalar2005} is given in Fig. \ref{fig:Protograph_Representation} together with its simplified representation.

%% file: Sections/PEXIT.tex
The protograph extrinsic information transfer (PEXIT) analysis is a method for easily determining the threshold of protograph-based LDPC codes \cite{Liva2007}. In the PEXIT analysis, we track the mutual information (MI) evaluation along the edges of the protograph during decoding. The MI between the code bits associated to (protograph) variable node $j$ and the messages computed by variable node $j$ sent towards check node $i$ is denoted $I_{\text{E}_{\text{v}}}^{(\ell)}(i,j)$ and computed using
        \begin{align}
        \label{PEXITb}
            I_{\text{E}_{\text{v}}}^{(\ell)}(i,j) = J\Bigg(&\Bigg[\sum_sb_{s,j}\bigg(J^{-1}\left(I_{\text{E}_{\text{c}}}^{(\ell-1)}(s,j)\right)\bigg)^2 -\\\nonumber &\!\!\!\!\!\bigg(J^{-1}\left(I_{\text{E}_{\text{c}}}^{(\ell-1)}(i,j)\right)\bigg)^2 + \bigg(J^{-1}\left(I_{\text{ch},j}\right)\bigg)^2\Bigg]^{\frac{1}{2}}\Bigg),
        \end{align}
where  $\ell$ denotes the current iteration of the PEXIT analysis.
\\\indent $J(\sigma)$ is the capacity of a binary input AWGN channel with noise variance $\sigma^2$:
\begin{equation}
    J(\sigma) = 1-\int_{-\infty}^\infty \frac{1}{\sqrt{2\pi\sigma^2}}e^{-\frac{(\tau - \frac{\sigma^2}{2})^2}{2\sigma^2}}\log_2(1+e^{-\tau})\differential\tau.
    \label{Eq:J}
\end{equation}
This $J$-function is often approximated using the approximations in \cite{TenBrink2004} or \cite{Brannstrom200}. Although these approximations are sufficiently accurate for high rate codes, these methods are not accurate enough for the optimization of ultra low rate protographs. Thus, we propose to use the Gauss-Hermite approximation\cite{Liu1994} and approximate
\begin{equation}
    J(\sigma) \approx 1 - \sum_{i=1}^\mu \frac{\alpha_i}{\sqrt{\pi}}\log_2\left(1+e^{-2\sigma x_i + \frac{\sigma^2}{2}}\right),
\end{equation}
where $\mu$ denotes the number of sample points, and $\alpha_i$ and $x_i$ are the weights and roots of the Hermite polynomial, respectively. We have found that $\mu = 100$ is sufficient for an accurate estimation of the threshold.
\\\indent In~\eqref{PEXITb}, $I_{\text{ch},j}$ is the mutual information between the channel input and output for variable node $j$, which for a given BI-AWGN channel characterized by ${E_\text{b}}/{N_0}$ is equal to $I_{\text{ch},j} = J(8R E_\text{b}/{N_0})$.
For a punctured variable node, $I_{\text{ch},j} = 0$ as this node is not transmitted over the channel.  
 $I_{\text{E}_{\text{c}}}^{(\ell)}(i,j)$ is the mutual information between the code bits and the messages computed by check node $i$ sent towards variable node $j$ and is given by
\begin{align}
  I_{\text{E}_{\text{c}}}^{(\ell)}(i,j) = 1-J\Bigg(\Bigg[\sum_{s}b_{i,s}\bigg(J^{-1}\left(1-I_{\text{E}_{\text{v}}}^{\ell}(i,s)\right)\bigg)^2 - \nonumber\\ \bigg(J^{-1}\left(1-I_{\text{E}_{\text{v}}}^{\ell}(i,j)\right)\bigg)^2\Bigg]^{\frac{1}{2}}\Bigg).
\end{align}
We say that a code converges if $I_{\text{APP}}^{\ell}(j) \to 1 \ \forall j$ as $\ell \to \infty$, where 
        \begin{align}
        \label{PEXITe}
            I_{\text{APP}}^{(\ell)}(j) = J\Bigg(\!\Bigg[\sum_s b_{s,j}&\bigg(J^{-1}\left(I_{\text{E}_{\text{c}}}^{l}(s,j)\right)\bigg)^2 + \nonumber\\  &\qquad\qquad\bigg(J^{-1}\left(I_{\text{ch},j}\right)\bigg)^2\Bigg]^{\frac{1}{2}}\Bigg)
        \end{align}
is the MI between the a posteriori probability at the decoder output and the respective code bit associated to variable node~$j$.

%% file: Sections/Protograph_Design.tex
\begin{figure}[t!]
    \centering
	\includegraphics[width=1\columnwidth]{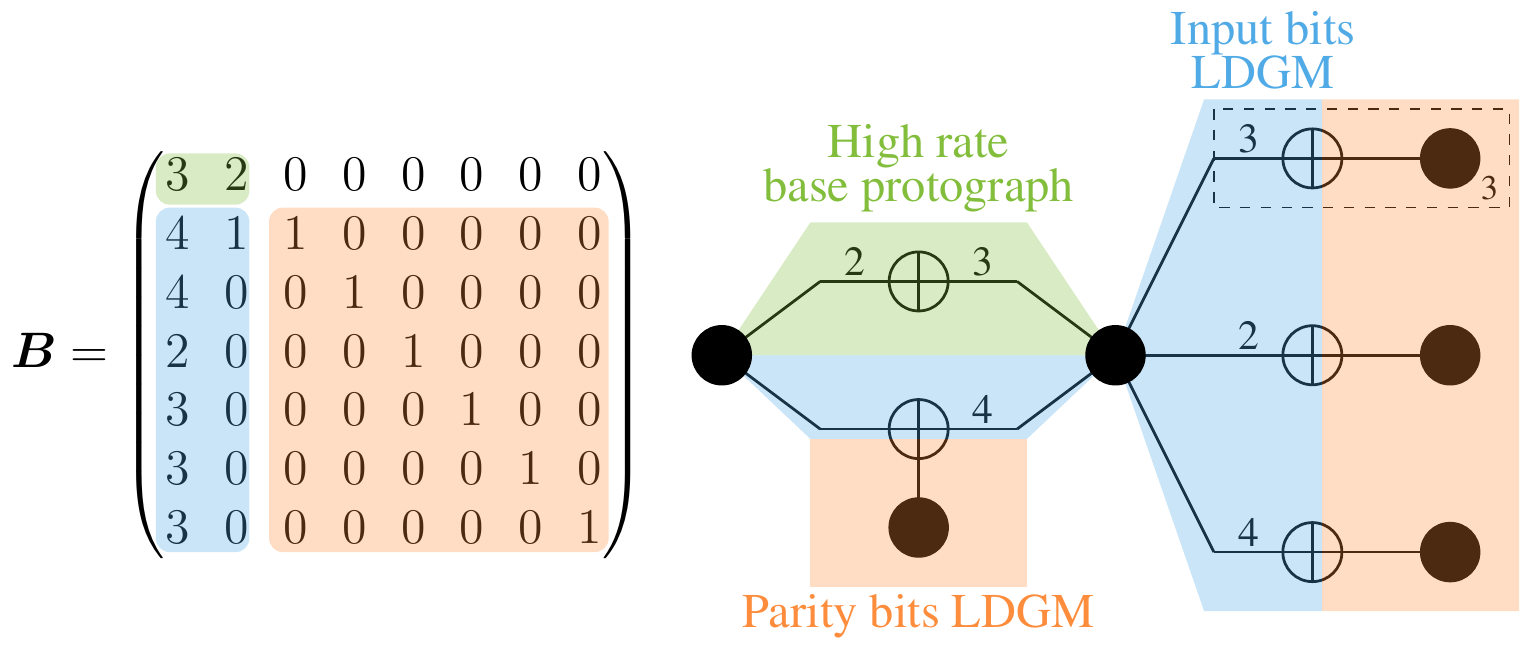}
	\vspace*{-3ex}
    \caption{Protograph structure for a rate $\frac{1}{8}$ code which consists out of the concatenation of a high rate protograph and an LDGM code.}
    \label{fig:Protograph_LDGM}
\end{figure}

Designing low rate protographs is more difficult compared to high rate protographs because the number of optimizable entries of the protomatrix grows quadratically as the rate of the code decreases. The search space of a protograph is determined by the size of the protomatrix and the maximum allowed value $e_\text{p}$ within the protomatrix, and is equal to $(1+e_\text{p})^{mn}$. Finding the optimum protograph using a brute force search quickly grows infeasible for low code rates.

One method of simplifying the design of a protograph is by fixing a part of the protograph, while carrying out the optimization only for the non-fixed parts \cite{Uchikawa2014}. This will decrease the search space of the optimization. One downside, however, is that the global optimum may be excluded from the search space by fixing parts of the protograph. It is therefore imperative that the fixed part is chosen such that a good performance results.

One design that has been shown to work well for low rate LDPC codes is the concatenation of a high rate dense protograph and an LDGM code~\cite{Jouguet2011},\cite{Garcia2003},\cite[Ch.~4]{Mani2021}. In LDGM codes, all check nodes in the code are connected to degree-one variable nodes, which correspond to the parity bits of the LDGM code. The remaining variable nodes correspond to the input bits of the LDGM code. Figure~\ref{fig:Protograph_LDGM} shows an example of the concatenation of an LDGM code with a high rate base protograph. Note that the variable nodes corresponding to the input bits of an LDGM code can also be of degree one. 

The particular protograph shown in Fig. \ref{fig:Protograph_LDGM} (rate $\frac{1}{8}$) was obtained using differential evolution \cite{Garcia2003,Uchikawa2014} by fixing the orange part (identity matrix) of the protomatrix, therefore optimizing the remaining green and blue parts. A similar optimization for different codes with rates ranging from $\frac{1}{5}$ to $\frac{1}{10}$ leads to very similar matrices with clusters of variable and check nodes of the same type. Here, we define a node type as a group of nodes wherein all the members have the same degree and therefore the same decoding convergence behaviour. A node type consists either solely of check nodes or variable nodes. Two variable (check) nodes have the same degree if they have the same number of connections to check (variable) nodes of the same type.

We illustrate the concept of nodes types using the example protograph of Fig.~\ref{fig:Protograph_LDGM}. The check nodes corresponding to the bottom 3 lines in $\boldsymbol{B}$ are grouped together and are of the same type. All three of these check nodes have three edge connections to the first variable node and are also connected to a different degree one variable node. These latter variable nodes are of the same type as they all have one connection to the aforementioned check nodes. Instead of optimizing every single entry in the protomatrix itself, we propose to specify a set of check node types, and optimize only the occurrence of each check node type. This reduces the search space of the protograph optimization significantly and renders the optimization for ultra low code rates feasible.

\begin{figure}[t!]
    \centering
	\includegraphics{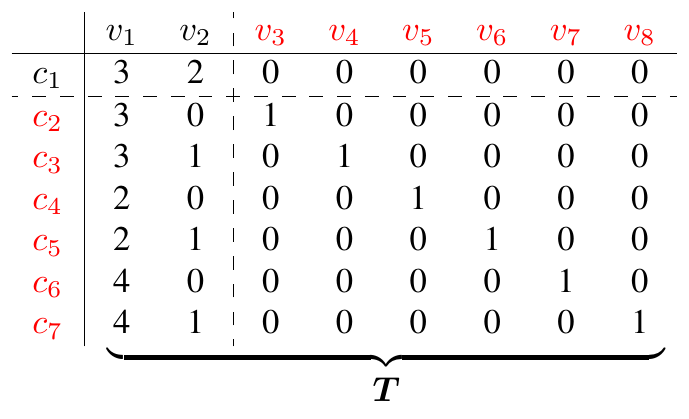}
	\vspace*{-2ex}
    \caption{Type description $\boldsymbol{T}$ for our suggested protomatrix optimization. In this case $K = 7$, $k = 1$, $L = 8$ and $l = 2$.}\vspace*{-2ex}   \label{fig:Protograph_Optimization}
\end{figure}
We propose the type description $\boldsymbol{T}$ shown in  Fig. \ref{fig:Protograph_Optimization} for optimizing low rate protographs. This type description corresponds to a concatenation of a high rate base protograph with an LDGM code. The rows represent the different check node types and the columns represent the different variable node types. $t_{i,j}$ is the value in position $(i,j)$ of the type description and represents the number of edge connections between a check node of type $i$ and a variable node of type $j$. In the example of Fig.~\ref{fig:Protograph_Optimization}, there are seven different check nodes types and eight different variable nodes types. 

We further subdivide the types into two classes, fixed node types and optimizable node types. We represent the node types as a set of $K$ check node types $\{\text{CN}_1, \ldots \text{CN}_k, \text{CN}_{k+1} \ldots \text{CN}_K\}$ and $L$ variable node types $\{\text{VN}_1, \ldots \text{VN}_l, \text{VN}_{k+1} \ldots \text{VN}_L\}$. Here, $k$ denotes the amount of fixed check node types and $l$ the amount of fixed variable node types. The vectors $\boldsymbol{c} = (c_1,\ldots c_K)$  and $\boldsymbol{v} = (v_1,\ldots v_L)$  specify how often check node type $\text{CN}_j$ occurs ($c_j$) and how often variable node type $\text{VN}_i$ occurs ($v_i$). The fixed node types only occur once within the protograph. Therefore $c_i = 1$ for $i < k$ and $v_j = 1$ for $j < l$.  During the optimization, we do not change the occurrence of the fixed node types, similar to how we normally fix parts of the protomatrix.

Optimizable check and variable node types are allowed to occur more than once in a protograph. In order to preserve the degrees of the optimizable node types, the occurrences of the check nodes types are linked to those of the variable node types. If an optimizable check node type CN$_i$ is connected to an optimizable variable node type VN$_j$, $c_i = v_j$ must hold. Therefore, the total number of optimizable check node type occurrences $h =  \sum_{i = k+1}^K c_j = m-k$ equals the total amount of optimizable variable node type occurrences.

The design rate of the code is determined by the total number of fixed node types, and the total number of occurrences of all optimizable check nodes types and is given by $ R = \frac{l - k}{l +h}$.
Note that a type description can be used to design codes of different rates by varying $\boldsymbol{c}$ and $\boldsymbol{v}$ (and hence $h$). 

\begin{figure}[t!]
    \centering
	\includegraphics[width=1\columnwidth]{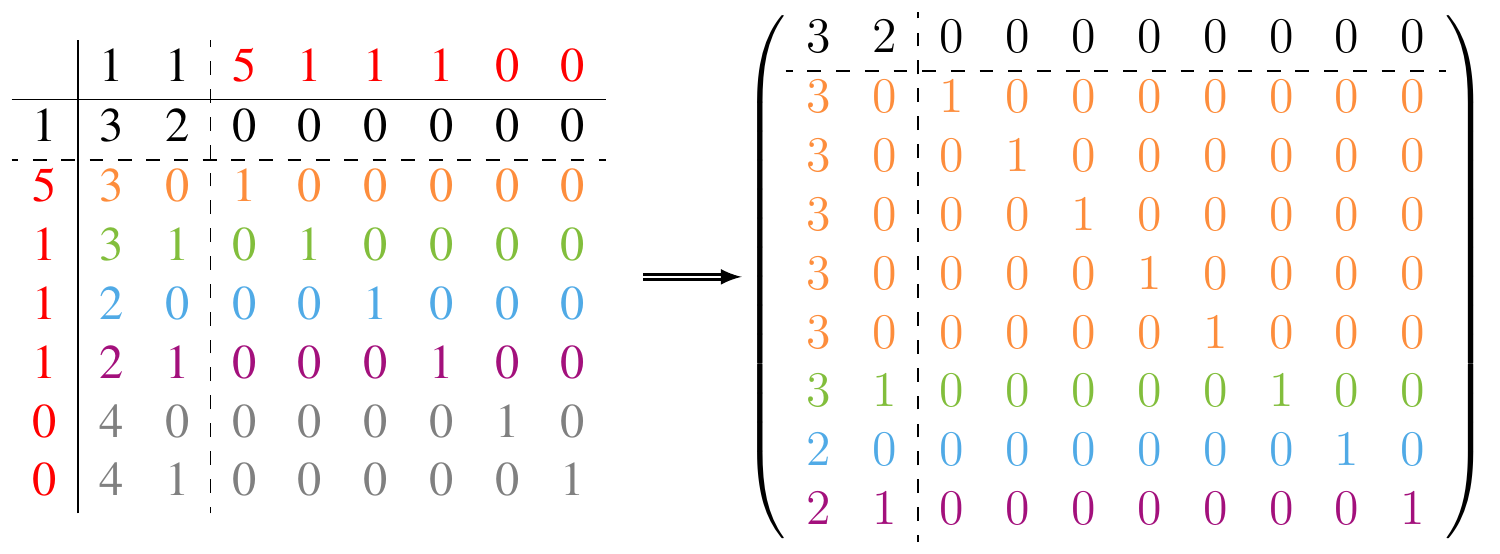}\vspace*{-2ex}
    \caption{Example of transforming the type description into a rate $\frac{1}{10}$ type-based protograph (TBP).}
    \label{fig:BasematrixToProtograph}
\end{figure}
Figure~\ref{fig:BasematrixToProtograph} shows an example of how to obtain a protomatrix from a type description. For better visibility, we color the optimizable check node types $\text{CN}_{k+1},\ldots \text{CN}_K$ using distinct colors. When constructing the protomatrix, its values have to be chosen such that the degrees of the optimizable nodes do not change. The degrees of the fixed nodes change depending on how often each optimizable nodes occurs. We use differential evolution similar to~\cite{Uchikawa2014} to optimize the non-fixed parts of $\boldsymbol{c}$ and $\boldsymbol{v}$ (marked in \textcolor{red}{red} in Figs.~\ref{fig:Protograph_Optimization}-\ref{fig:Protograph_Expansion})\footnote{Source code carrying out the optimization can be found online at the repository https://github.com/kadirgumus/Protograph\_Optimization}. During the optimization, we need to make sure that $h$ stays consistent in order to fulfill the rate constraint of the code. 
 \begin{figure}[t!]
    \centering
     \resizebox{\linewidth}{!}{\def\herespace{\,\,\,\,\,}
\def\herespacex{\,\,}
\def\colw{3ex}
\newcolumntype{x}{%
>{\centering\hspace{0pt}}p{2ex}}%
\begin{tabular}{@{}c|c@{\herespace}c@{\herespace}c@{\herespace}c@{\herespace}c@{\herespace}c : c@{\herespace}c@{\herespace}c@{\,\,\,}c@{\herespacex}c@{\herespacex}c@{\herespacex}c@{\herespacex}c@{\herespacex}c@{\herespacex}c@{}}
  & $v_{1}$ &	 $v_{2}$ &  $v_{3}$&  $v_{4}$&  $v_{5}$ & $v_{6}$ & \textcolor{red}{$v_{7}$} & \textcolor{red}{$v_{8}$} & \textcolor{red}{$v_{9}$} & \textcolor{red}{$v_{10}$} & \textcolor{red}{$v_{11}$} & \textcolor{red}{$v_{12}$} & \textcolor{red}{$v_{13}$} & \textcolor{red}{$v_{14}$} & \textcolor{red}{$v_{15}$} & \textcolor{red}{$v_{16}$}\\\hline
$c_{1}$  & 1 & 1 & 1 & 1 & 1 & 0 & 0 & 0 & 0 & 0 & 0 & 0 & 0 & 0 & 0 & 0\\ 
$c_{2}$  & 1 & 1 & 1 & 1 & 0 & 1 & 0 & 0 & 0 & 0 & 0 & 0 & 0 & 0 & 0 & 0\\ 
$c_{3}$  & 1 & 1 & 1 & 0 & 1 & 1 & 0 & 0 & 0 & 0 & 0 & 0 & 0 & 0 & 0 & 0\\\hdashline 
\textcolor{red}{$c_{4}$}  & 1 & 1 & 1 & 0 & 0 & 0 & 1 & 0 & 0 & 0 & 0 & 0 & 0 & 0 & 0 & 0\\ 
\textcolor{red}{$c_{5}$}  & 1 & 1 & 0 & 1 & 0 & 0 & 0 & 1 & 0 & 0 & 0 & 0 & 0 & 0 & 0 & 0\\ 
\textcolor{red}{$c_{6}$}  & 1 & 1 & 0 & 0 & 1 & 0 & 0 & 0 & 1 & 0 & 0 & 0 & 0 & 0 & 0 & 0\\ 
\textcolor{red}{$c_{7}$}  & 1 & 1 & 0 & 0 & 0 & 1 & 0 & 0 & 0 & 1 & 0 & 0 & 0 & 0 & 0 & 0\\ 
\textcolor{red}{$c_{8}$}  & 1 & 0 & 1 & 1 & 0 & 0 & 0 & 0 & 0 & 0 & 1 & 0 & 0 & 0 & 0 & 0\\ 
\textcolor{red}{$c_{9}$}  & 1 & 0 & 1 & 0 & 1 & 0 & 0 & 0 & 0 & 0 & 0 & 1 & 0 & 0 & 0 & 0\\ 
\textcolor{red}{$c_{10}$}  & 1 & 0 & 1 & 0 & 0 & 1 & 0 & 0 & 0 & 0 & 0 & 0 & 1 & 0 & 0 & 0\\ 
\textcolor{red}{$c_{11}$}  & 0 & 1 & 1 & 1 & 0 & 0 & 0 & 0 & 0 & 0 & 0 & 0 & 0 & 1 & 0 & 0\\ 
\textcolor{red}{$c_{12}$}  & 0 & 1 & 1 & 0 & 1 & 0 & 0 & 0 & 0 & 0 & 0 & 0 & 0 & 0 & 1 & 0\\ 
\textcolor{red}{$c_{13}$}  & 0 & 1 & 1 & 0 & 0 & 1 & 0 & 0 & 0 & 0 & 0 & 0 & 0 & 0 & 0 & 1\\ 
\end{tabular}}
    \caption{Example of an expanded type description with $\tilde{q} = 4$.}\vspace*{-2ex}
    \label{fig:Protograph_Expansion}
\end{figure}

Using this method, the size of the search space reduces to ${S + h-1 \choose h-1}$, where $S = K - k$ is the number of different optimizable check node types. For $S\geq 2$, we can (coarsely) upper-bound  ${S + h-1 \choose h-1}$ by $S^h$. 
If $S < (1+e_\text{p})^n$, then, because $h \leq m$, the search space for the TBP optimization is smaller than that of the conventional optimization as $S^h < (1+e_\text{p})^{mn}$. If the largest value allowed in the protomatrix is $e_\text{p}$, then there are \mbox{$(1+e_\text{p})^n$} possible ways to fill in each row, i.e., there are \mbox{$(1+e_\text{p})^n$} different possible check node types. As we only utilize a select few check node types in the proposed type description, $S < (1+e_\text{p})^n$. 
For example, assuming that we only optimize the input bits of the LDGM part of the protomatrix as shown in Fig.~\ref{fig:Protograph_LDGM} (light blue part) when optimizing a rate $\frac{1}{10}$ code with the type description given in Fig. \ref{fig:Protograph_Optimization}, we have $m^\prime = 8$ non-fixed check nodes, $n^\prime = 2$ non-fixed variable nodes, $S = 6$ and $h = 8$.
The amount of different such protographs equals $1716$, while the search space size equals $1.5\cdot10^{11}$ for the method in~\cite{Uchikawa2014}. For the optimization of lower rate protographs, this gap grows even larger.

\begin{figure*}[t!]
    \centering
	\includegraphics[width=\textwidth]{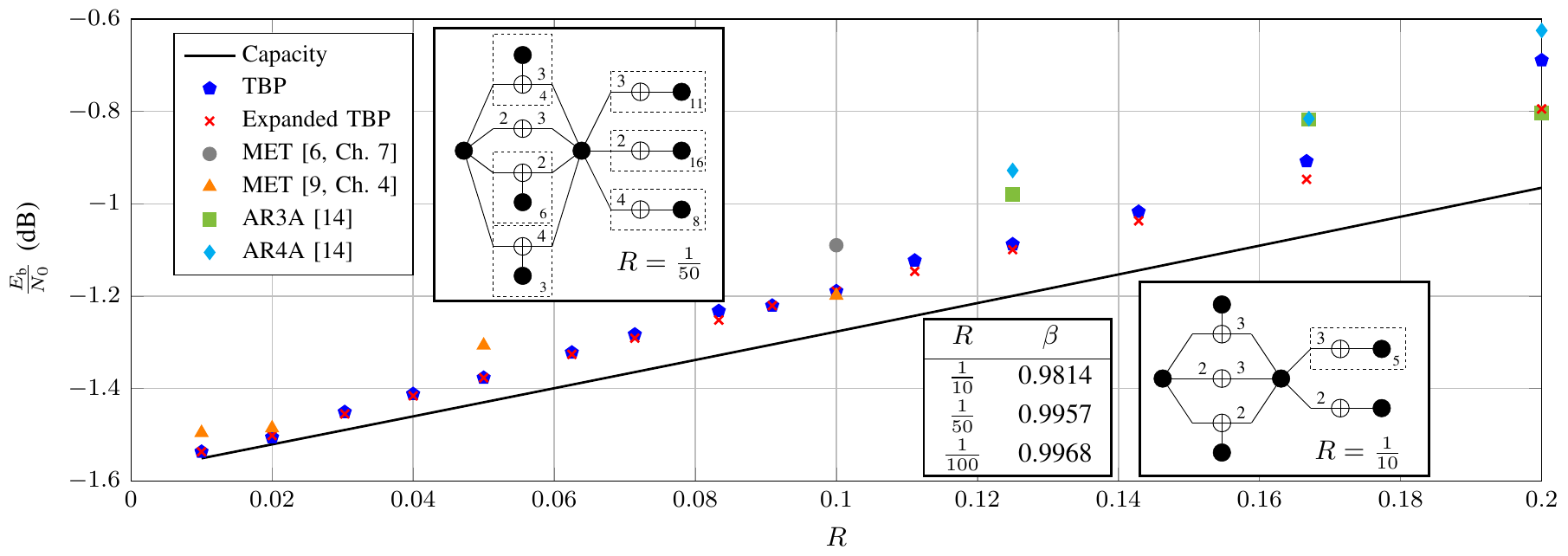}
	\vspace*{-5ex}
    \caption{The thresholds of codes optimized for a variety of different code rates for both the TBP and the lifted TBP ($\tilde{q} = 4$). The codes from the literature are taken from \cite{Divsalar2005,Mani2021}.}
    \label{fig:Optimization_thresholds}
\end{figure*}

For codes of rates $\frac{1}{6}$ and above, the search space using our proposed type-based approach is too restricted. To increase the search space, we propose to use an expanded binary ($e_{\text{p}} = 1$) type description obtained from lifted versions (lifting factor~$\tilde{q}$) of the original type description. We then use the expanded type description to obtain a larger protomatrix of size $\tilde{q}m\times \tilde{q}n$. With a large $\tilde{q}$, the protomatrix can be fine-tuned more, but the complexity of the optimization increases. In this paper we choose $\tilde{q} = 4$, such that the optimization is still feasible. An example of an expanded type description is given in Fig.~\ref{fig:Protograph_Expansion}.

%% file: Sections/Modified_PEXIT.tex
\indent For our proposed optimization, the PEXIT analysis can be simplified, as check nodes and variables nodes of the same type have the exact same convergence behaviour. So instead of calculating the PEXIT functions for each variable node and check node, we only have to carry out the calculation for each check and variable node type. We can replace $b_{i,j}$ in (\ref{PEXITb})--(\ref{PEXITe}) by $t_{i,j}$. We need to adjust the calculation of $I_{\text{E}_\text{v}}(i,j)$ for the fixed variable node types as
        \begin{align}
            I_{\text{E}_\text{v}}^{(\ell)}(i,j) = J&\Bigg(\Bigg[\sum_st_{s,j}c_s\bigg(J^{-1}\left(I_{\text{E}_\text{c}}^{(\ell - 1)}(s,j)\right)\bigg)^2 - \\\nonumber &\bigg(J^{-1}\left(I_{\text{E}_\text{c}}^{(\ell - 1)}(i,j)\right)\bigg)^2 + \bigg(J^{-1}\left(I_{\text{ch},j}\right)\bigg)^2\Bigg]^{\frac{1}{2}}\Bigg),
        \end{align}
        and $I_{\text{APP}}$ for the fixed variable node types as
                \begin{align}
            I_{\text{APP}}^{(\ell)}(j) = J\Bigg(\Bigg[\sum_s t_{s,j}c_s\bigg(J^{-1}\left(I_{\text{E}_\text{c}}^{(\ell -1)}(s,j)\right)\bigg)^2 + \\\nonumber \bigg(J^{-1}\left(I_{\text{ch},j}\right)\bigg)^2\Bigg]^{\frac{1}{2}}\Bigg).
        \end{align} This speeds up the threshold computation, as the complexity grows only with the size of $\boldsymbol{T}$ instead of the size of $\boldsymbol{B}$.

%% file: Sections/Results.tex
\begin{figure*}[t!]
    \centering
	\includegraphics[width=1\textwidth]{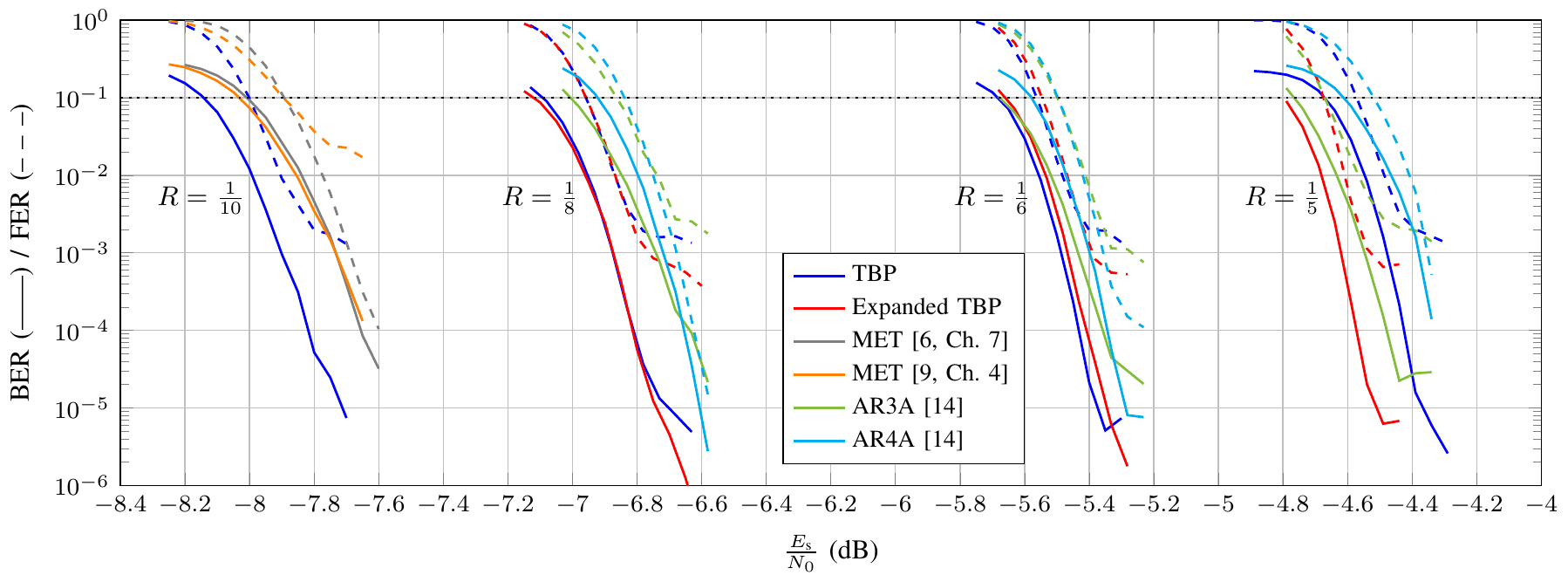}
	    \vspace*{-5ex}
    \caption{FER (dashed lines, --\ --\ --) and BER (solid lines, ------) simulation results of several different protographs of different rates as a function of $E_{\text{s}}/N_0 = R\cdot E_{\text{b}}/N _0$.  The codes from the literature are taken from \cite{Richardson2002,Divsalar2005,Mani2021}. All codes have a blocklength $N \approx 10^5$.}\vspace*{-3ex}
    \label{fig:Protograph_FER}
\end{figure*}

In order to verify the effectiveness of our proposed TBP optimization, we optimized protographs with both the standard and expanded TBP ($\tilde{q} = 4$) for a wide variety of code rates. For the standard TBP optimization, we use the type description in Fig. \ref{fig:Protograph_Optimization} for all code rates. For the expanded TBP, we use an expanded type description\footnote{The type description for the expanded TBP optimization can be found online at {https://github.com/kadirgumus/Protograph\_Optimization}} based on the one from Fig.~\ref{fig:Protograph_Optimization}. Figure~\ref{fig:Optimization_thresholds} shows the thresholds of the codes that we have obtained and compares it to both the capacity of the BI-AWGN channel and some codes from the literature. The insets show the optimized protographs for the rate $R=\frac{1}{10}$ and rate $R=\frac{1}{50}$ codes in the simplified representation. Figure \ref{fig:Optimization_thresholds} also contains a table with the reconciliation efficiencies for different codes. The codes created with the TBP optimization method outperform or are just as good as the codes in the literature for all code rates. As the rate of the codes decreases, the gap to capacity decreases as well, which suggests that this particular design works quite well for ultra low rate codes. As mentioned before, the TBP optimization method works less well for higher rate codes due to the restricted search space. Using expanded TBPs leads, however, to a threshold improvement for high rate codes. For low rates, the expansion does not yield substantial threshold gains.

In order to verify the performance of our protographs, we have carried out finite-length simulations and measured FER and BER curves for several codes and compare them to the literature. All of the parity check matrices for these codes were generated randomly, after which we removed 4-cycles such that the matrix fulfills the row-column constraint. The length of the codewords $N=nq$ is approximately equal to $N = 10^5$ for the rate $\frac{1}{5}$ to $\frac{1}{10}$ codes and $N= 10^6$ for the rate $\frac{1}{50}$ and $\frac{1}{100}$ codes. We assume transmission over a BI-AWGN channel and use sum-product decoding with 500 decoding iterations.

\begin{figure}[t!]
    \centering
	\includegraphics[width=1\columnwidth]{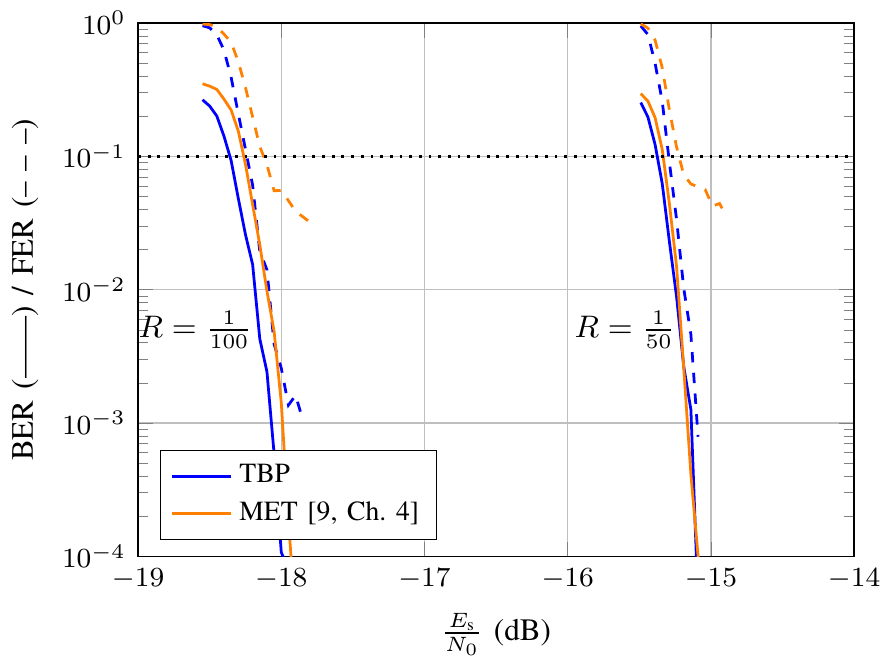}
	 \vspace*{-5ex}
    \caption{FER (dashed lines, --\ --\ --) and BER (solid lines, ------) simulation results of several different protographs of different rates as a function of $E_{\text{s}}/N_0 = R\cdot E_{\text{b}}/N _0$. All codes have a blocklength $N \approx 10^6$.}\vspace*{-3ex}
    \label{fig:Protograph_FER2}
\end{figure}

Figure \ref{fig:Protograph_FER} and \ref{fig:Protograph_FER2} show the result of these simulations. We can see that for all code rates, our TBP codes outperform the codes in the literature by up to 0.12\,dB at an FER of 0.1, except for $R = \frac{1}{5}$, where our code is as good as the reference. Our codes also show a clear error floor behaviour, where the height of the error floor decreases as the code rate decreases. Note that we did not take into account the error floor behaviour during both the protograph optimization and the construction of the parity-check matrices. Further note that the FER simulations in \cite[Ch.~4]{Mani2021} for the $R = \frac{1}{50}$ code show no such error floor. The parity check matrix for this code used in the simulations in \cite[Ch.~4]{Mani2021}, however, was designed using progressive edge growth, which generally lowers the error floor.

%% file: Sections/Conclusion.tex
In this paper, we have presented TBPs, a new method for optimizing and representing protograph-based LDPC codes for the ultra low rate regime. Using differential evolution, we have obtained codes that beat the state-of-the-art over a wide range of code rates. We have verified the performance of our codes using Monte-Carlo simulations and find that they outperform the reference codes in the regime that is particularly useful for CV-QKD. We have furthermore proposed the expanded TBP method, which remedies some drawbacks when designing higher rate codes. Possible future work includes the further extension of this method towards the high-rate code regime and further optimization of the type description.